# Surface-passivated high-$Q$ GaAs photonic crystal nanocavity with quantum dots

Kazuhiro Kuruma[1, a)], Yasutomo Ota[2, b)], Masahiro Kakuda[2], Satoshi Iwamoto[1,2] and Yasuhiko Arakawa[2]

[1]*Institute of Industrial Science, The University of Tokyo, 4-6-1 Komaba, Meguro-ku, Tokyo 153-8505, Japan*
[2]*Institute of Nano Quantum Information Electronics, The University of Tokyo, 4-6-1 Komaba, Meguro-ku, Tokyo 153-8505, Japan*
a) E-mail: kuruma@iis.u-tokyo.ac.jp
b) E-mail: ota@iis.u-tokyo.ac.jp

**Abstract**
**Photonic crystal (PhC) nanocavities with high quality ($Q$) factors have attracted much attention because of their strong spatial and temporal light confinement capability. The resulting enhanced light-matter interactions are beneficial for diverse photonic applications, ranging from on-chip optical communications to sensing. However, currently achievable $Q$ factors for active PhC nanocavities, which embed active emitters inside, are much lower than those of the passive structures because of large optical loss, presumably originating from light scattering by structural imperfections and/or optical absorptions. Here, we demonstrate a significant improvement of $Q$ factors up to ~160,000 in GaAs active PhC nanocavities using a sulfur-based surface passivation technique. This value is the highest ever reported for any active PhC nanocavities with semiconductor quantum dots. The surface-passivated cavities also exhibit reduced variation in both $Q$ factors and cavity resonant wavelengths. We find that the improvement in the cavity performance presumably arises from suppressed light absorption at the surface of the PhC's host material by performing a set of PL measurements in spectral and time domains. With the surface passivation technique, we also demonstrate a strongly-coupled single quantum dot-cavity system based on a PhC nanocavity with a high $Q$ factor of ~100,000. These results will pave the way for advanced quantum dot-based cavity quantum electrodynamics and for GaAs micro/nanophotonic applications containing active emitters.**

## I. INTRODUCTION

Semiconductor optical micro/nanocavities that can strongly confine photons in time and space are promising platforms for research on light-matter interactions and for developing various photonic devices, including lasers[1,2], quantum light sources[3,4], and sensors[5,6]. Among reported optical cavities, photonic crystal (PhC) nanocavities have been intensively studied because of their high quality ($Q$) factors with small mode volumes ($V$) on the order of one cubic optical wavelength. These properties are advantageous for enhancing light-matter interactions in cavity quantum electrodynamics (CQED) systems. In particular, III-V semiconductor-based PhC nanocavities have been widely used for active nanophotonic devices and for coupling with solid-state quantum emitters, such as self-assembled quantum dots (QDs). Two dimensional GaAs-based PhCs are one of the leading platforms among them, and have realized high-$Q$ PhC nanocavities coupled with single QDs in a strong coupling regime[7], together with a variety of fascinating CQED phenomena[8–12]. Pursuit of stronger coupling with high-$Q$ cavities is important for improving diverse optical devices, including nanocavity lasers[13,14], ultrafast optical switches[15–17], non-classical light generators[18,19], and quantum logic gates[20]. At this stage, a further increase in $Q$ factor while maintaining high QD-cavity coupling strength is expected for further enabling advanced CQED devices that are inaccessible with current device performances[21,22].

In general, an experimental $Q$ factor ($Q_{exp}$) is lower than the one in design because of structural imperfections and optical absorptions. It is known that design $Q$ factors ($Q_{des}$s) can be enormously large in several dedicated PhC structures, such as those with precise airhole shifts around cavity region[23], leading to $Q_{des}$s of $10^5 \sim 10^9$[24,25]. Such high $Q$ factors have been experimentally confirmed in passive PhC nanocavities fabricated in various dielectrics/semiconductors such as Si[26] and GaAs[27]. Recently, suppression of surface absorptions was found to play a key role in further improvement of $Q_{exp}$ by surface treatments[28,29]. So far, surface passivation techniques for mitigating the surface absorption have been investigated for various semiconductor optical micro/nanocavities, including those using atomic layer deposition[29,30], chemical treatments[28,30,31], and gas adsorption[32]. Meanwhile, reported $Q_{exp}$ values for active PhC nanocavities till date still exhibit larger discrepancies between $Q_{des}$ and $Q_{exp}$. The latter is often limited to the order of $10^4$[17,33,34], even when the PhC structure is precisely patterned. This suggests significant influence of optical absorptions, such as those in active emitters and host materials, and at the surface of the cavities. However, the influence of such optical absorption on $Q_{exp}$ in active PhC nanocavities is still unclear.

Indeed, there is no report on experimental approach to significantly improve $Q_{exp}$ for active PhC nanocavities.

In this study, we demonstrate a significant increase in $Q_{exp}$ of GaAs active PhC nanocavities using a surface passivation technique. We employed a sulfur-based solution to modify the surface of the PhCs, which strongly reduces surface absorption and improves average $Q_{exp}$ by three times than that in the case without passivation. The best $Q_{exp}$ reaches ~160,000, which is the highest value among any active PhC nanocavities with QDs ever reported. Importantly, we have achieved the high $Q$ factors at around 900 nm, where the state-of-the-art In(Ga)As/GaAs QDs are available for coupling to the cavity mode. In addition, the surface passivation also enables large reduction of variations in $Q_{exp}$ and cavity resonance wavelength. We find that these improvements stem from the suppression of the surface nonradiative recombination process in GaAs after the treatment. We consistently observed enhanced PL emission and extended carrier lifetime of the GaAs bulk. The prolonged carrier lifetime was also confirmed as distinct behaviors in pump power dependences of $Q_{exp}$s and resonant wavelengths of cavities with and without passivation. We also demonstrate strong coupling with a single QD and passivated nanocavity with an exceedingly high $Q_{exp}$ of 100,000, paving the way for QD-based nanophotonic devices and PhC-based cavity QED systems operating with previously-inaccessible high $Q$ factors.

**II. CAVITY STRUCTURE AND SAMPLE FABRICATION**

The nanocavity investigated in this work is formed in an airbridge PhC with a double-heterostructure[35], as shown schematically in Fig. 1(a). The PhC consists of a triangular lattice of airholes in the GaAs slab with a line-defect waveguide formed by a missing row of airholes along Γ–K direction. We choose the lattice constant, airhole radius, and slab thickness to be $a_1$ = 252 nm, $r$ = 61 nm, and $d$ = 130 nm, respectively. The double-hetero region is defined by slightly enlarging the lattice constant ($a_2$) to 259.6 nm in Γ-K direction, which forms an optical mode-gap and gently confines photons inside [35]. This cavity structure is advantageous for achieving high experimental $Q$ factors[26]. $Q_{des}$ for the cavity design is $1\times10^7$ for the fundamental cavity mode, calculated by using the 3D finite difference time domain (FDTD) method. We also introduced double-periodic modulation of airhole radii by ±1.2% around the cavity region for better extraction of cavity radiation at the expense of the reduction in $Q_{des}$[36]. We finally obtained a $Q_{des}$ of $5\times10^5$ and a mode volume of $1.5(\lambda/n)^3$, where $\lambda$ (= $3.875a$) is the cavity resonant wavelength and $n$ (= 3.46) is the refractive index of GaAs. Figure 1(b) shows a calculated electric field distribution of the investigated fundamental cavity mode.

The designed PhC cavity was fabricated into a 130-nm-thick GaAs slab using electron beam (EB) lithography combined with both dry and wet etching. First, the PhC patterns were written into a resist layer by EB exposure and then transferred onto a GaAs slab using a dry etching process. The airbridge GaAs slab structure was finally formed by removing a 1-µm-thick $Al_{0.7}Ga_{0.3}As$ sacrificial layer underneath the slab using hydrofluoric acid. The GaAs slab contains a single layer of InAs QDs at the middle. We chose a sufficiently low QD density (~$10^8$ cm$^{-2}$) to prevent unnecessary optical absorption in the active layer. PL peaks of single QDs were mainly observed around 930 nm, which were used for strong coupling with a PhC cavity in later experiments. Figure 1(c) shows a scanning electron microscope (SEM) image of a fabricated cavity taken after the wet etching process. A cross sectional SEM picture of a similar PhC can be found in ref.[37].

From the SEM images of the cavity, we evaluated variations in airhole radii and positions in the fabricated cavities to quantify the influence of the structural disorders on $Q_{exp}$. We fitted the airhole edges in the images with circles using ProSEM software (GenISys Corp.). The inset of Fig. 1(d) shows a fitting result. Here, we used test PhC cavities without the ±1.2% radius modulations for simplifying the evaluation process. Figure 1 (d) shows a summary of the extracted radii from 623 different airholes. From fitting with a Gaussian function, we deduced average radius and standard deviation to be Avg.($r$) = 61.1 nm and S.D.($r$) = 0.4 nm, respectively. The obtained Avg.($r$) shows a good agreement with that designed of 61 nm. Imperfections of airhole positions were also evaluated using the same software; the standard deviations of the $x$ and $y$ positions are 0.69 and 0.64 nm, respectively. Next, we used these extracted values for estimating the impact of the structural disorders on $Q$ factors (denoted as $Q_{diso}$) by FDTD simulations[38]. In the estimation, we assumed that the influence of sidewall roughness of airholes on $Q$ factors is much smaller than that of variations in airhole radii and positions[39]. We evaluated 30 different patterns of cavities with the same radius and position fluctuations evaluated experimentally. The computed average and standard deviation of $1/Q_{diso}$ are Avg.($1/Q_{diso}$) = 2.6×10$^{-6}$ and S.D.($1/Q_{diso}$) = 5.8×10$^{-7}$, respectively. The corresponding $Q_{diso}$ still maintains a high value of ~400,000, exhibiting a discrepancy with $Q_{exp}$s, as discussed later. This suggests the need to consider the influence of optical absorptions on $Q_{exp}$.

In order to reduce the surface absorptions of the fabricated nanocavities, we performed sulfur-based surface passivation using a $Na_2S$ solution. Optical absorption at the surface of GaAs could be associated with surface defect states presumably induced by surface oxidation[40], which could form shallow light-absorption energy levels inside the GaAs bandgap. The sulfur-based surface passivation based on $Na_2S$ solution is known to eliminate the oxide layer on the GaAs surface. Simultaneously, sulfur atoms bond with surface-exposed GaAs atoms[41,42], which leads to the reduction of the surface states in the bandgap, resulting in suppressed light absorption. Figures 2 (a)–(e) outline the basic flow of the investigated passivation process. After removing the sacrificial layer by wet etching to form the PhC slab structure (Fig. 2(a)), we performed wet chemical digital etching[43] for surface cleaning by the removal of native oxide. For this step, we first prepared a hydrogen peroxide ($H_2O_2$) solution ($H_2O_2 : H_2O = 1 : 9$) and dipped the samples for 30 seconds to oxidize them (Fig. 2(b)). Then, the samples were rinsed with pure water and subsequently etched by a 1 mol/L solution of citric acid ($C_6H_8O_7$) for 1 min (Fig. 2(c)). After rinsing the samples, we kept them in pure water to prevent oxidation in air. Next, we dissolved $Na_2S \cdot 9H_2O$ in isopropyl alcohol (IPA, 20 ml) at 170 °C and cooled it down to room temperature to make a supersaturated $Na_2S$ solution. For passivation, we immersed the samples into the $Na_2S$ solution for 10 minutes (Fig. 2(d)). We note that the use of IPA is beneficial for increasing the efficiency of passivation, since the low electric permittivity of IPA (~20 at 20 °C) leads to an increase in the electrostatic interaction of sulfide ions with surface atoms of GaAs[42,44]. After the passivation, the samples were rinsed with deionized water and loaded into a helium-flow cryostat in a glove box under a $N_2$ atmosphere to avoid sample oxidation (Fig. 2(d)). The sample loading process was conducted after the oxygen concentration inside the glove box becomes <1000 ppm. For a fair comparison with passivated sample, we prepared another sample without the $Na_2S$ treatment by dividing one sample chip into two pieces after the wet digital etching process.

### III. OPTICAL CHARACTERIZATION
**A. Statistical Evaluation of $Q_{exp}$ and $\lambda$**

In order to evaluate $Q_{exp}$s of the fabricated PhC cavities, we performed reflectance measurements using super luminescent diode (SLD) at 60 K (see supplementary material for details of the measurement setup). For extracting $Q_{exp}$, the measured cavity spectra were fitted by a Voight function while fixing its Gaussian part to be our spectrometer response (the details of fitting procedure can be found in supplementary material). For the following analyses, we also employ the inverse of $Q$ factor, $1/Q$, which indicates optical loss. Figure 3 (a) shows histograms of measured $1/Q_{exp}$ for 60 different samples with (red) and without (light green) passivation, respectively. Obviously, overall $1/Q_{exp}$ for the samples with passivation are smaller than those without passivation, demonstrating the improvement in $Q_{exp}$. By fitting the histograms with a Gaussian function, the average $1/Q$ values (Avg.($1/Q_{exp}$)) for the samples with and without passivation were deduced to be $1.16\times10^{-5}$ and $2.98\times10^{-5}$, respectively. The corresponding Avg.($Q_{exp}$) for the passivated samples is 90,000, which is roughly three times higher than those without passivation—35,000. In order to quantify the reduction in optical absorption loss, we used the following equation:

$$\frac{1}{Q_{exp}} = \frac{1}{Q_{des}} + \frac{1}{Q_{str}} + \frac{1}{Q_{abs}}, \qquad (1)$$

where $Q_{des}$, $Q_{str}$, and $Q_{abs}$ are $Q$ factors respectively associated with design, structural imperfections, and optical absorptions. $Q_{des}$ ($1/Q_{des}$) is $5\times10^5$ ($2\times10^{-6}$) according to the FDTD simulations, while $Q_{str}$ ($1/Q_{str}$) is deduced to be $1.71\times10^6$ ($5.85\times10^{-7}$) using simulation results for PhC cavities with measured structural fluctuations as stated above. We also assume that the values of $Q_{des}$ and $Q_{str}$ are common for the passivated and unpassivated samples. Under these assumptions, the measured three-fold increase in $Q_{exp}$ can be interpreted as a 67% reduction in the optical absorption loss: passivation reduced $1/Q_{abs}$ from $2.73\times10^{-5}$ to $9.03\times10^{-6}$. These results highlight the impact of the passivation process on increasing $Q$ factor. We also found that the statistical variations of $1/Q_{exp}$ values for the passivated samples are smaller than that for samples without passivation, as seen in Fig. 3(a). The S.D. of $1/Q_{exp}$ (S.D.($1/Q_{exp}$)) for the passivated and unpassivated samples are $2.36\times10^{-6}$ and $6.81\times10^{-6}$, respectively. This result suggests that the absorption loss before passivation is not uniform across the samples. For applications, lower variance is essential in, for example, increasing device fabrication success yield. The remaining variation in $1/Q_{exp}$ after the passivation is considered as originating from the remaining absorption loss, because the absorption loss still dominates $Q_{exp}$ in our nanocavities.

Figure 3 (b) shows extracted cavity wavelengths ($\lambda$s) for the samples with and without passivation. $\lambda$s for passivated samples become shorter than those for the unpassivated samples. The average $\lambda$ (Avg.($\lambda$)) for the passivated and unpassivated samples are found to be 946.4 nm and 954.8 nm, respectively. The decrease in Avg.($\lambda$) by 8.4 nm can be understood as a consequence of the homogeneous etching of the GaAs slab and airholes by Na$_2$S solution[45]. Corresponding etching depth was estimated to be 1.9 nm using FDTD simulations. We also found that the standard deviation of $\lambda$ (S.D.($\lambda$)) was reduced from 1.34 nm to 0.83 nm by the passivation, suggesting the usefulness of sulfur-based surface passivation in reducing the $\lambda$ variation. Compared to calculated variations, S.D.($\lambda$) = 0.19 nm and S.D.($1/Q_{diso}$) = 5.8×10$^{-7}$ using simulations with the measured structural fluctuations; we concluded that these variations in $\lambda$ and $1/Q_{exp}$ after the passivation do not arise from structural disorders of airhole radii and positions, again indicating that the remaining variations predominantly originate from remaining optical absorptions.

Figures 3 (c) and (d) respectively show the cavity reflectance spectra with the highest $Q_{exp}$s among the samples with and without passivation. While the extracted $Q_{exp}$ for the unpassivated sample was ~47,000, we obtained a 3-times higher $Q_{exp}$ of ~160,000 for the sample with passivation, which is, to the best of our knowledge, the highest value ever reported for any active PhC nanocavities embedding semiconductor QDs. We note that the sulfur-based surface passivation prevents GaAs surface from being oxidized by air[46]. Indeed, we did not see significant degradation of $Q_{exp}$s for the passivated samples even after the cavities were exposed to air for two days in total. Moreover, we confirmed that our passivation technique can be used for largely improving $Q_{exp}$s of other nanocavity structures such as so-called L3 and H0 types.

**B. Evaluation of the GaAs Surface based on PL Measurements**

For investigating the cause of reduced absorption loss by the passivation, we performed photoluminescence (PL) measurements of the GaAs slab in the unpatterned region (details of the PL measurement can be found in supplementary material). Figure 4(a) shows the PL spectrum of GaAs band edge measured for samples with (red) and without (black) passivation at room temperature. The peak of spectrum around 870 nm originates from the GaAs band edge. We observed a 1.5-fold increase in PL intensity after the passivation. Figure 4 (b) exhibits time-resolved PL decay curves of the GaAs band edge measured for samples with (red curve) and without (black curve) passivation, plotted together with fitting curves with a double-exponential function. The observed non-single exponential decay is attributed to a result of high carrier injection and to a large contribution from nonlinear carrier recombination such as Auger processes[47]. Of the deduced two decay time constants, the slower component was attributed as reflecting carrier lifetime. The carrier lifetime for the passivated sample was deduced to be 230 ps, which is 1.6 times longer than that for the unpassivated sample. From the measured lifetime of the passivated sample, we estimated the surface recombination velocity $S$ of the GaAs surface to be 6×10$^4$ cm/s, which is comparable with the previously reported value for a passivated sample with Na$_2$S[41]. In the estimation of $S$, we assumed the $S$ of the interface between the GaAs slab and AlGaAs sacrificial layer to be negligible (~500 cm/s)[47]. The prolonged carrier lifetime is consistent with the increased PL intensity and suggests the reduction in nonradiative surface recombination centers by the passivation.

Next, we examined the behavior of $Q_{exp}$s under the presence of free carriers to further verify the extended carrier lifetime in passivated PhC cavities. In this experiment, we injected free carriers into the GaAs barrier via above-band-gap excitation using an 808 nm continuous wave laser diode, and measured $Q_{exp}$ through cavity reflectance measurements with varying power of the carrier injection laser. Figure 4 (c) shows the dependence of the $Q_{exp}$ on excitation powers ranging from 10 nW to $4.5 \times 10^7$ nW. At low excitation powers, both samples exhibit constant $Q_{exp}$s (indicated as black and blue dashed lines) since only negligible amounts of carriers are injected into the GaAs slab. With further increasing excitation powers, both $Q_{exp}$s start degrading because of free carrier absorption but with different onset powers. The passivated sample degrades even from a weak excitation power of around 200 nW. In contrast, the unpassivated sample degrades from a much higher excitation power of around $10^5$ nW. The onset of the degradation is determined by the balance between carrier injection and recombination[48]. The passivated sample accumulates carriers more efficiently because of the enhanced carrier lifetime, and thus, degrades earlier than the unpassivated sample.

Marked influences of prolonged carrier lifetime were also observed in cavity resonant wavelength ($\lambda$), as shown in Fig. 4(d). $\lambda$s for the unpassivated sample does not exhibit any significant changes until a pump power of $10^7$ nW. In contrast, $\lambda$s for the passivated sample are gradually blueshifted with increasing pump power, suggesting larger carrier plasma effect[49] induced by higher carrier density. At an excitation power of $10^7$ nW, we observed the largest blueshift of -0.125 nm for the passivated sample, as indicated by the black arrow in Fig 4 (b). The measured blueshift can be explained by a simple model considering a typical expression of carrier plasma effect, including the measured $S$ (see supplementary material). Again, the observed blueshift of $\lambda$ is consistent with earlier onset of the strong degradation of $Q_{exp}$, which can be explained by prolonged carrier lifetime. For higher excitation powers over $10^7$ nW, we observed rapid redshifts for both samples because of heating[48,50]. At the highest excitation power of $4.5 \times 10^7$ nW, the unpassivated cavity showed the largest redshift of 0.28 nm for the sample without passivation, which corresponds to temperature rise of the sample to ~7 K, estimated from the measured relationship between sample temperature and shift of the $\lambda$ (see supplementary material).

## IV. STRONG COUPLING BETWEEN A SURFACE-PASSIVATED CAVITY AND SINGLE QD

Finally, we performed PL spectroscopy to demonstrate strong coupling between a single QD and a high-$Q$ PhC nanocavity realized by the surface passivation. For this study, we used a slightly different PhC cavity design with lattice constants of $a_1$ = 249 nm and $a_2$ = 256.5 nm for coupling to QDs emitting around 930 nm. Figure 5 (a) shows a PL spectrum of a surface-passivated nanocavity taken under a far QD-cavity detuning condition at 9.6 K. By fitting the spectrum, we deduced a high cavity $Q$ factor of ~100,000 (corresponding to a cavity decay rate $\kappa$ of 13 μeV). This is the highest value among any PhC nanocavity ever used for QD-CQED study. For evaluating the intrinsic $Q$ factor of the cavity, we used sufficiently low excitation power of 23 nW to avoid the carrier-induced degradation of cavity $Q$. We noticed that PL signals from most cavities we measured increase after the passivation, as expected from the prolonged carrier lifetime in the system. This property is highly beneficial for practical applications using active PhC devices driven by carrier injection.

We then investigated the properties of optical coupling between the cavity mode and a QD exciton emission line at 934.01 nm. For control of the QD-cavity spectral detuning, we tuned the cavity wavelength using a Xe gas condensation technique[51], which did not deteriorate the $Q_{exp}$ even after repeating the tuning process several times. Figure 5 (b) shows the color map of the PL spectra when the cavity wavelength is tuned across the QD emission line by the gas deposition. We observed an anti-crossing between the QD and cavity mode around the QD-cavity resonance condition, indicating that the QD-cavity system is in a strong coupling regime. The inset in Fig. 5(b) shows a vacuum Rabi spectrum at QD-cavity resonance condition, overlaid with fitting curves (blue solid lines). The additional center peak between two polariton peaks originates from a bare cavity emission[8,52–54], which is turned on by blinking of the QD and is supplied by pumping from the QD in off-resonant states and/or background emission[55]. Figure 5(c) shows a summary of extracted peak positions of polariton branches and the bare cavity mode by fitting with multi-Voigt functions, again exhibiting the anti-crossing of the system. From the separation between two polariton branches at QD-cavity resonance, we deduced a vacuum Rabi splitting of 40 μeV, which corresponds to a coupling strength $g$ of 20 μeV. The measured peak positions can be well reproduced by using a standard CQED model[8], as plotted in Fig. 5(c) using black solid lines. Owing to the very small $\kappa$, the QD-cavity coupled system still fulfills the strong coupling condition ($g > \kappa/4$), despite the relatively small $g$. Such a strongly-coupled QD-cavity system with small $g$ and low $\kappa$ is beneficial for observing clear vacuum Rabi oscillations under limited time resolutions of the time domain measurements[33]. These results also demonstrate the strongly-coupled QD-CQED system using a surface-passivated PhC cavity with an exceedingly high $Q$ factor to the order of $10^5$.

## V. CONCLUSION

In conclusion, we have demonstrated large improvements in $Q$ factors of GaAs active PhC nanocavities using sulfur-based surface passivation. The passivation technique largely reduced optical absorption, leading to a three-fold improvement of average $Q$ factor than that without passivation. For the best cavity, we realized a high $Q$ factor of ~160,000, the highest value for any active PhC nanocavities embedding QDs. In addition, we observed reduced variations in $Q$ factors and cavity wavelengths after the passivation. We conclude that these improvements in the passivated cavities mainly originate from the reduction in surface recombination centers of GaAs, which was supported by observing the increase in PL intensity and carrier lifetime. The enhanced carrier lifetime was also confirmed by measuring excitation power dependences of $Q$ factors and cavity wavelengths. Finally, we demonstrated strong coupling between a single QD and a passivated PhC cavity with a high $Q$ factor of ~100,000. Our results show the importance of suppressing surface absorptions for achieving high experimental $Q$ factors as well as for reducing variations in $Q$ factors and cavity wavelengths among different samples. Our study will be a great help not only for further exploring advanced CQED experiments that are inaccessible with the current QD-cavity coupled systems [56,57], but also for diverse GaAs micro/nanophotonic devices requiring high-$Q$ active cavities.

## SUPPLEMENTARY MATERIAL

See the supplementary material for the details regarding (a) the optical measurement setup, (b) fitting of the cavity spectra for evaluating $Q$ factors, (c) estimation of cavity wavelength shifts due to carrier plasma effect, and (d) temperature dependence of cavity wavelengths for evaluating the rise of sample temperature caused by laser heating.

## ACKNOWLEDGMENTS

We would like to thank M. Nishioka and S. Ishida for their technical support and A. Osada for fruitful discussions. This work was supported by JSPS KAKENHI Grant-in-Aid for Specially Promoted Research (15H05700), KAKENHI (16K06294, 18J13565, 19K05300), JST PRESTO (JPMJPR1863), Inamori Foundation, Iketani Science and Technology Foundation, Murata Science Foundation and a project of the New Energy and Industrial Technology Development Organization (NEDO).

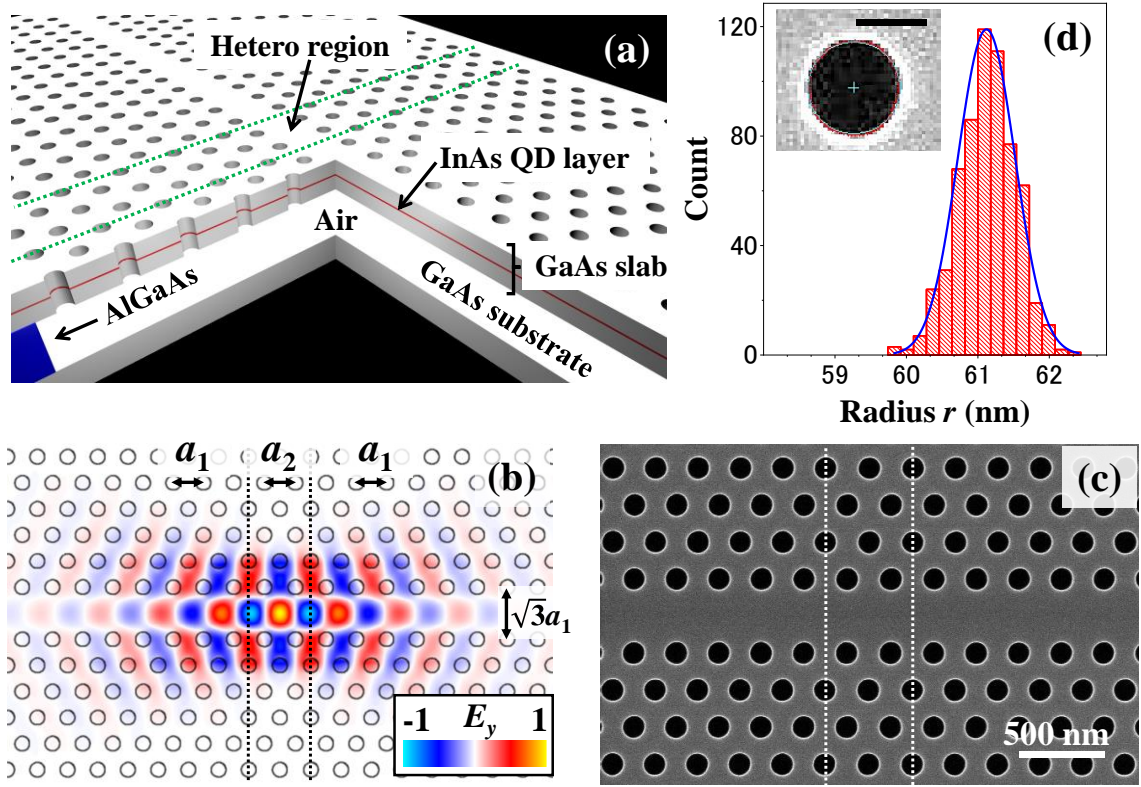

Fig. 1. (a) Schematic of a PhC double-heterostructure cavity based on an air-suspended structure. A single InAs QD layer is buried in the middle of the GaAs slab. The green dashed lines indicate interfaces between regular PhCs and the hetero region. (b) Electric field ($E_y$) profile of the investigated fundamental cavity. The interfaces between the regular PhCs with a lattice constant of $a_1$ = 252 nm and the hetero region with $a_2$ = 259.6 nm are shown as black-dashed lines. (c) SEM image of a fabricated cavity. The white-dashed lines correspond to the hetero interfaces. (d) Histogram of the extracted airhole radii ($r$) of 623 airholes. The blue solid line is a fitting curve with a Gaussian function. The average and standard deviation of $r$ is 61.1 nm and 0.4 nm, respectively. The inset shows a SEM image of an airhole in a PhC nanocavity. The blue circle is a fitting curve to the airhole edge (red data points). The blue cross mark indicates the deduced center of the airhole. The black bar indicates a 100 nm scale bar.

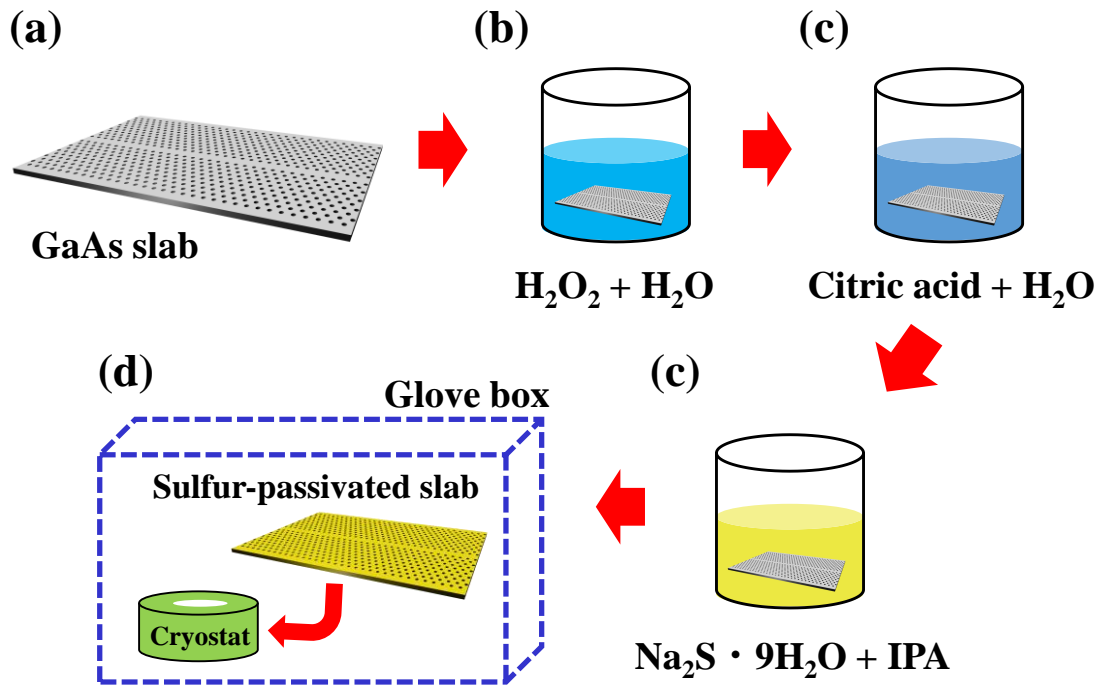

Fig. 2. Process flow of the investigated, sulfur-based surface passivation. The process starts with preparing (a) an airbridge PhC cavity. A digital etching process for surface cleaning by dipping into a (b) $H_2O$-diluted $H_2O_2$ solution for the oxidation of the cavity surface before being etched by a (c) $H_2O$-diluted citric acid solution. Subsequently, the cavity is (c) immersed into a supersaturated $Na_2S$ solution based on isopropyl alcohol (IPA) and then (d) loaded into a cryostat inside a glove box under $N_2$ atmosphere to prevent surface oxidation.

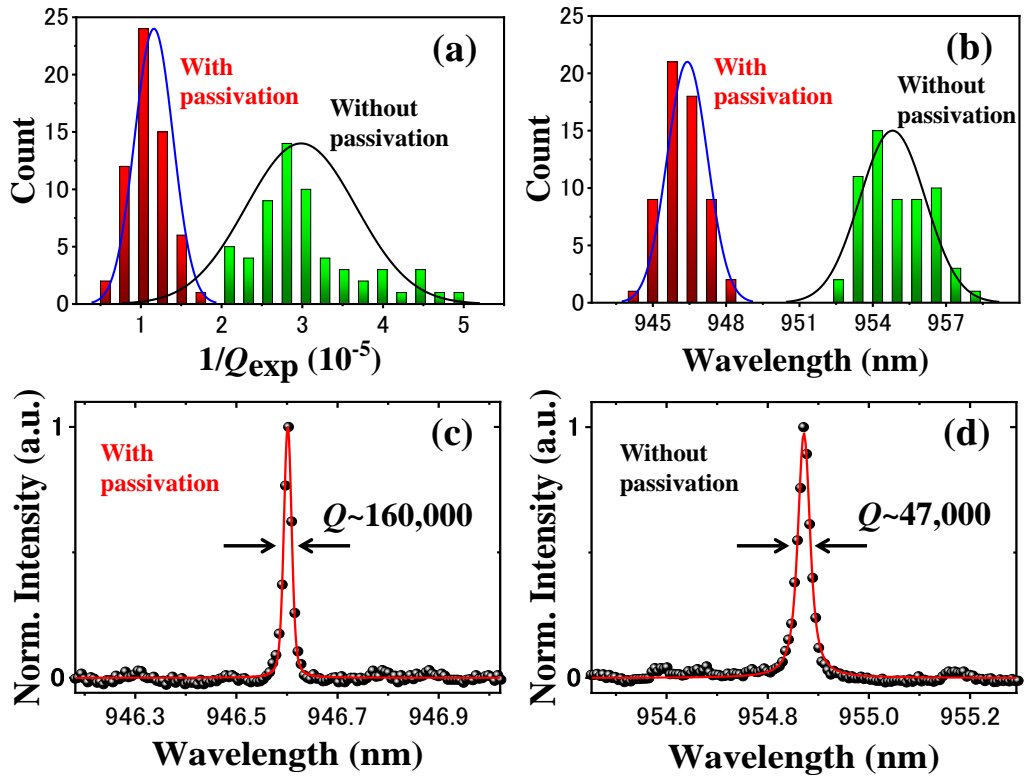

Fig. 3. Histograms of (a) reciprocal experimental $Q$ factors ($1/Q_{exp}$s) and (b) cavity wavelengths measured for 60 different samples with (red) and without (light green) passivation, respectively. The blue and black solid lines respectively represent fitting results using a Gaussian function for the two samples series. (c,d) Selected cavity reflection spectra with the highest $Q_{exp}$ among the measured samples (c) with passivation and (d) without passivation. The red solid lines indicate fitting curves using a Voigt peak function when fixing its Gaussian linewidth to be the spectrometer response.

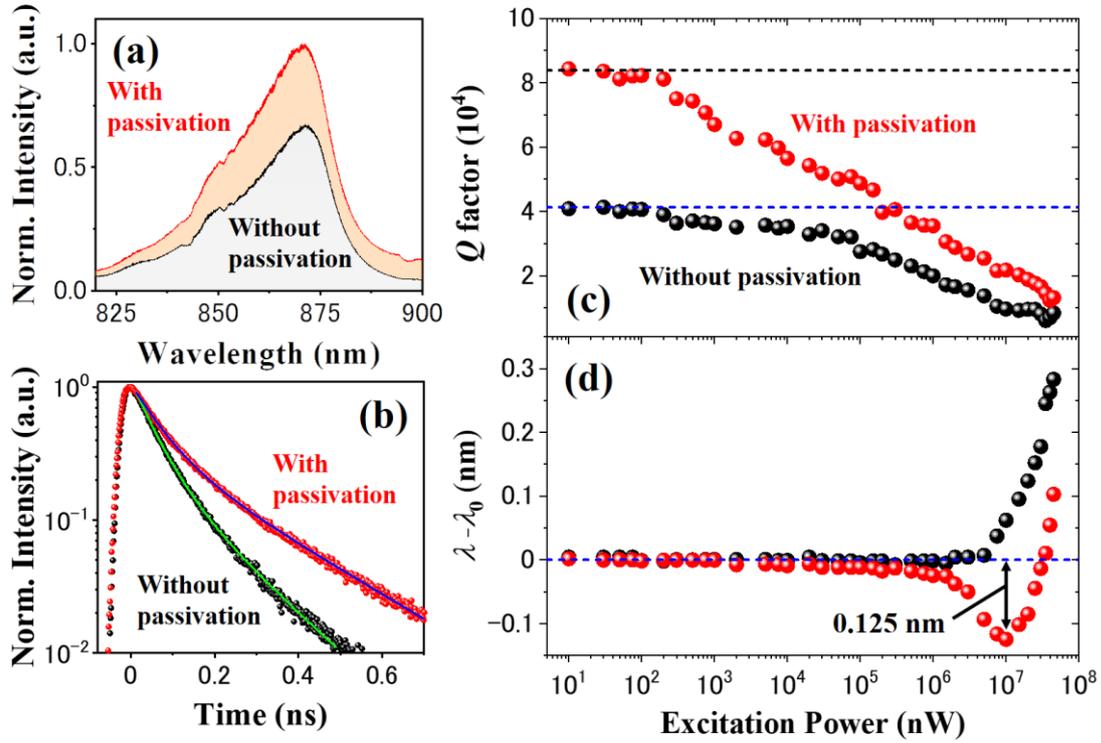

Fig. 4. (a) Spectra of emission near the GaAs band edge measured for the samples with (red) and without (black) passivation taken with an average excitation power of 500 μW at room temperature. (b) Corresponding PL decay curves to the spectra presented in (a). The blue and light green lines are fitting results with double-exponential function. (c) Experimental $Q$ factors as a function of excitation power of the 808-nm carrier injection laser. The red and black points respectively indicate experimental data measured for a cavity with and without passivation. The black (blue) dashed line indicates the initial value measured under zero carrier injection for the sample with (without) passivation. (d) Corresponding shifts of cavity wavelengths ($\lambda$) to the samples discussed in (c). The shifts are plotted with respect to the resonance wavelength measured at 0 nW ($\lambda_0$). The blue dashed line shows the zero shift.

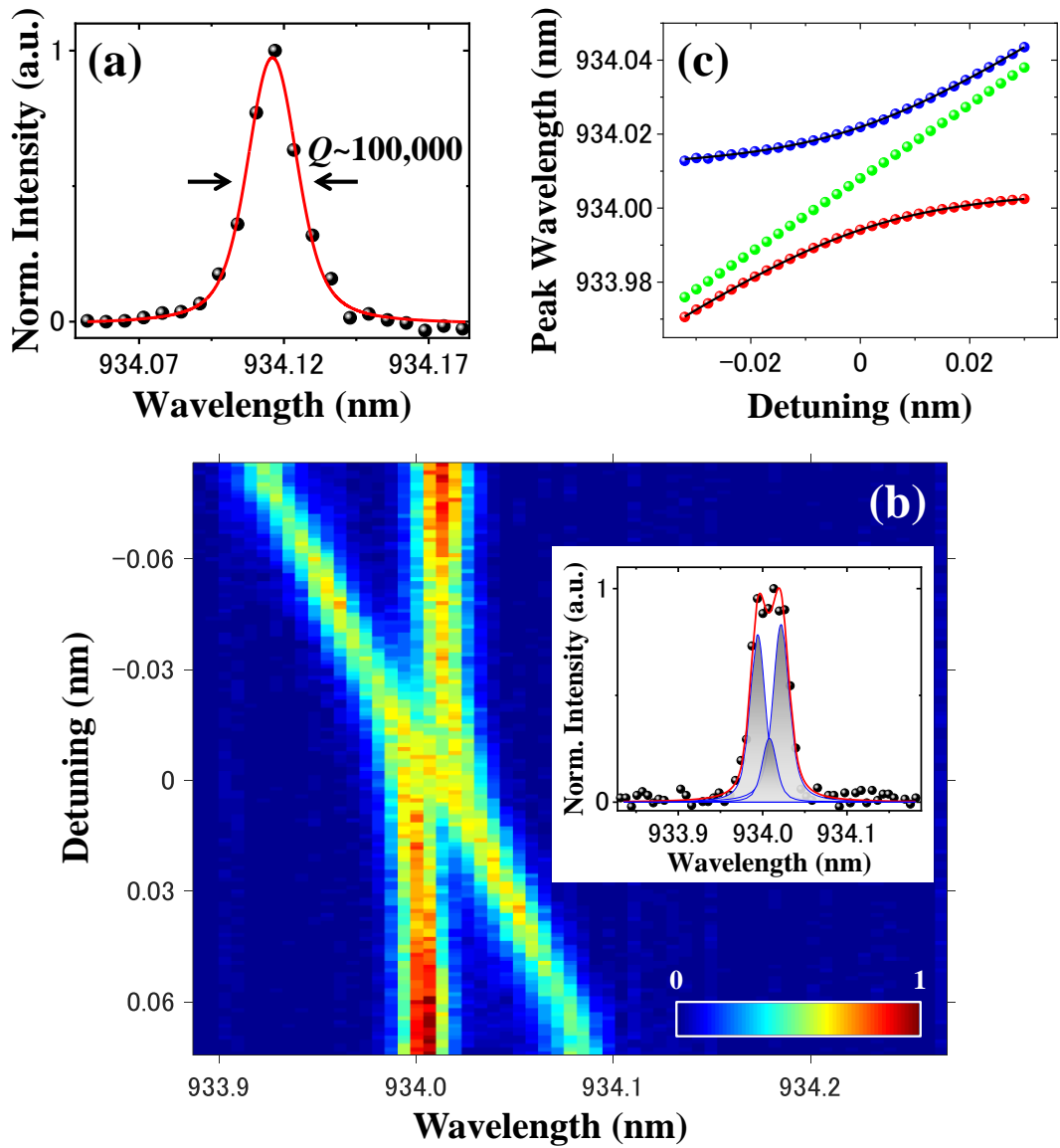

Fig. 5. (a) PL spectrum of the fundamental cavity mode measured for a sample under a far-detuned condition. The red solid line is a fitting curve. (b) Color map of PL spectra showing the cavity mode coming across the QD emission peak. The inset shows a vacuum Rabi spectrum under the QD-cavity resonance condition. The solid red line is a fitting result by multiple Voigt peak functions. The solid blue lines exhibit extracted peaks of the two polariton branches. The PL spectra were taken at an excitation power of 160 nW. (c) Peak wavelengths of the two polariton branches and the bared cavity mode extracted by the fitting. The blue, red, and light green dots denote the upper polariton, lower polariton, and bare cavity mode, respectively. The back solid lines are theoretically calculated dispersion curves.

# Supplementary Information for

# "Surface-passivated high-Q GaAs photonic crystal nanocavity with quantum dots"

Kazuhiro Kuruma[1], Yasutomo Ota[2], Masahiro Kakuda[2], Satoshi. Iwamoto[1,2] and Yasuhiko Arakawa[1,2]

[1]*Institute of Industrial Science, The University of Tokyo, 4-6-1 Komaba, Meguro-ku, Tokyo 153-8505, Japan*
[2]*Institute of Nano Quantum Information Electronics, The University of Tokyo, 4-6-1 Komaba, Meguro-ku, Tokyo 153-8505, Japan*

## 1. Optical measurement setup

A simplified schematic of the optical measurement set up is shown in Fig. S1. We kept the PhC samples in a continuous-flow liquid-helium cryostat equipped with a temperature controller. For reflectance measurements, we used a super luminescent diode (SLD) with broadband emission ranging from 900 nm–1050 nm. For above-band-gap excitation of GaAs, we employed two different lasers. The first one is a Ti:sapphire pulse laser (pulse duration:1 ps, repetition rate: 80 MHz and center wavelength: 780 nm) and was used for the optical characterization of GaAs band edge emission. The second one is an 808 nm continuous wave (CW) diode laser and was used for measuring strong coupling in QD-cavity systems. The same laser was used for characterizing pump power dependences of $Q$ factor and cavity wavelength of PhC nanocavities while monitoring cavity reflection spectra using the SLD. We focused all excitation light sources onto PhC cavities by an objective lens (OL) with a magnification factor of 50 and a numerical aperture of 0.65. The excitation power was measured before the OL. The optical power of SLD is fixed at 4.5 $\mu$W in the reflectance measurements. Signals from the samples were analyzed by a spectrometer equipped with a Si charge-coupled device (CCD) camera. In the reflectance measurements, we used two linear polarizers, which are set to be in front of and behind the center beam splitter (BS) to control the polarization of SLD light and to extract cavity signals from reflected light. After fixing the polarization of SLD by the front polarizer, we tuned the angle of the other polarizer to maximize the signal-to-noise ratio. In order to avoid measuring asymmetric cavity spectra induced by Fano interference, careful adjustment of the polarizer angles was necessary. For characterizing emission from GaAs bulk and QD-cavity coupled systems, we performed PL measurements after removing the two polarizers. For time-resolved PL measurements on the GaAs band edge, we used a time-correlated single-photon counting (TCSPC, Becker & Hickl Corp.) system equipped with a fast-response superconducting single-photon detector (SSPD, SCONTEL Corp.). The total time resolution of the system is measured to be 25ps. In the optical coupling experiment between a cavity and a QD, we utilized a Xe gas condensation technique[1] for cavity wavelength tuning. The same gas port employed for the Xe injection was used to introduce air into the cryostat for investigating the stability of *Q* factors under an ambient condition.

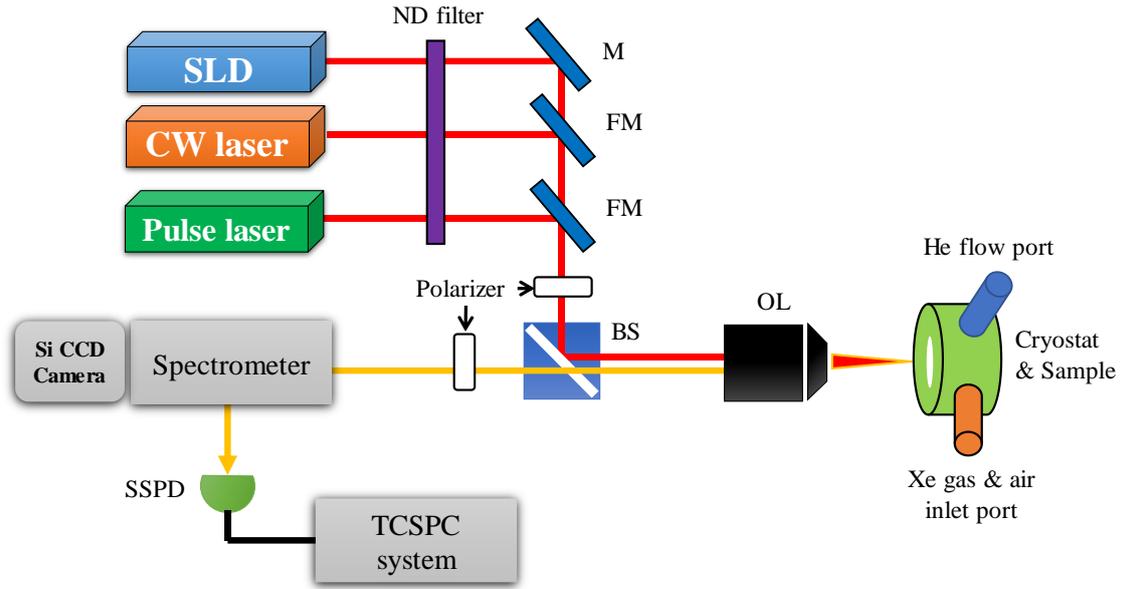

**Fig. S1. Illustration of the optical measurement setup. SLD: super luminescent diode; M: mirror; FM: flip mirror; OL: objective lens; BS: beam splitter.**

## 2. Experimental evaluation of *Q* factors

For the extraction of cavity *Q* factor, we fitted the cavity spectrum using a Voigt function. For the PL measurement of strong coupling in the QD-cavity system, we evaluated a bare cavity *Q* factor from a spectrum taken under a far-detuned condition. For the fitting with Voigt functions, we consider our spectrometer response by fixing the Gaussian linewidth to be the measured spectrometer response function of 21 $\mu$eV. From this fitting procedure, we can resolve a Lorentzian linewidth of the Voigt function as an intrinsic cavity linewidth with an accuracy far beyond the spectrometer response. In general, true linewidths can be extracted from convolved functions using an accurately-estimated instrumental response as long as they are well described with a mathematical model, as commonly used in, for example, high-resolution X-ray spectroscopy[2]. Empirically, this approach in our setup allows for accurate evaluation of $Q$ factor down to ~10 $\mu$eV.

### 3. Estimation of cavity wavelength shift due to carrier plasma effect

Here, we discuss the observed blueshift of cavity wavelength in Fig. 4(d) of the main text. The shift originates from the change of the refractive index $\Delta n$ because of carrier plasma effect. $\Delta n$ in carrier plasma effect is given by the following equation[3]:

$$\Delta n = -\frac{e^2 \lambda^2}{8\pi^2 c^2 \varepsilon_0 n_{GaAs}} \left( \frac{N_e}{m_e} + \frac{N_h}{m_h} \right) \tag{S1}$$

where, $c$, $\varepsilon_0$, and $e$ are light speed, permittivity of vacuum, and elementary charge, respectively; $m_e$ and $m_h$ are the effective masses of an electron and a hole, respectively. $n_{GaAs}$ is the refractive index of the GaAs slab (=3.46), and $\lambda$ is the wavelength of cavity mode. Here, we assume the $\Delta n$ is uniform across the whole cavity structure and the carrier density $N$ of the electron and hole are equal ($N_e \sim N_h$). For the calculation of $N$, we assumed typical physical properties of GaAs in its absorption coefficient[4] and carrier recombination processes[5,6] using the measured surface recombination lifetime of 230 ps and a laser spot size of 4 μm. For considering nonradiative surface recombination, we employed a simple expression for a surface recombination rate, $1/\tau_{nr} = S/d$, where $S$ is a surface recombination velocity. $d$ is in the dimension of length and can be represented by a slab thickness[7]. In this work, we assumed an effective slab thickness of 65 nm in order for taking into account the presence of air holes' sidewalls. Then, $N$ at an excitation power of $10^7$ nW can be calculated to be $2\times10^{17}$ cm$^{-3}$ with a conventional rate equation[7]. This value results in $\Delta n$ of -4×10$^{-4}$. Using the 3D FDTD simulations, the refractive index change results in $\Delta \lambda$ of -0.114 nm. The latter value is consistent with that observed in Fig. 4(d) in the main text. The difference between the theoretical and experimental values is probably because of the uncertainty in the physical properties of GaAs and the laser spot size used in the estimation of *N*.

**4. Temperature dependence of cavity wavelength**

In order to evaluate the rise of sample temperature caused by laser heating (shown in the pump power dependence of cavity wavelength ($\lambda$) in Fig. 4(d) of the main text), we investigated temperature dependence of $\lambda$ for a passivated sample. Figure S2 shows a dependence of $\lambda$ for temperature (*T*) from 60 to 80 K. *y* axis shows the difference between $\lambda$ and that measured at 60 K ($\Delta\lambda = \lambda - \lambda_{60K}$). We observed a linear increase of $\lambda$ as sample temperature increased. We fit the experimental data using the least square fitting (red solid line). From the fitting result, we deduced the temperature dependence of $\Delta\lambda$ to be 0.043 nm/K. From this value, we deduced the temperature increase at a pump power of $4.5\times10^7$ nW in Fig. 4(d) to be ~7 K.

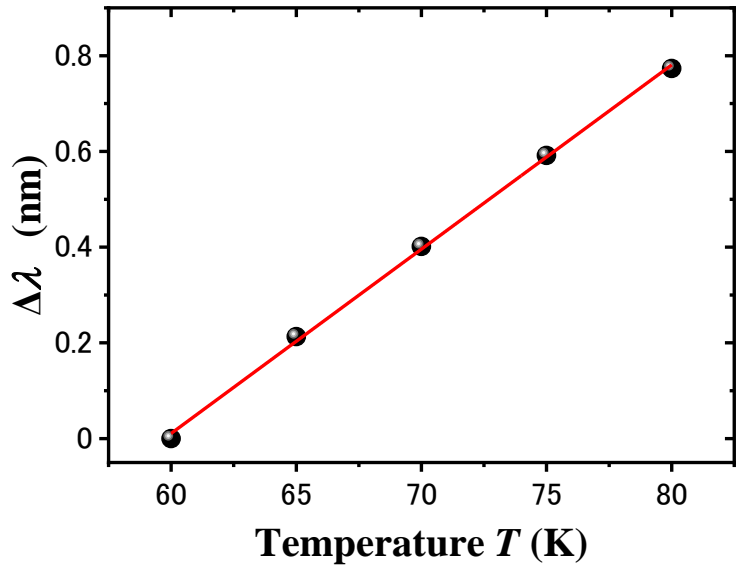

**Fig. S2. Difference between cavity wavelength from that measured at 60 K ($\Delta\lambda = \lambda - \lambda_{60K}$) plotted as a function of sample temperature (*T*). Red solid line is a linear fitting curve.**